\documentclass[11pt,journal,draftcls,letterpaper,onecolumn,oneside]{IEEEtran}

\IEEEoverridecommandlockouts                              
\usepackage{epsfig,latexsym,amsfonts,amsmath,amssymb,verbatim,cite,mathrsfs}
\usepackage{graphicx}
\usepackage[dvips]{color}

\def\rR{{\mathbb R}}
\def\eE{{\mathbb E}}
\def\pP{{\mathbb P}}

\def\@begintheorem#1#2{\tmpitemindent\itemindent\topsep 0pt\rm\trivlist
    \item[\hskip \labelsep{\indent\it #1\ #2:}]\itemindent\tmpitemindent}
\def\@opargbegintheorem#1#2#3{\tmpitemindent\itemindent\topsep 0pt\rm \trivlist
    \item[\hskip\labelsep{\indent\it #1\ #2\
    \rm(#3):}]\itemindent\tmpitemindent}
\def\@endtheorem{\endtrivlist\unskip}

\newtheorem{definition}{Definition}[section]

\newtheorem{theorem}{Theorem}[section]

\newtheorem{corollary}{Corollary}[section]
\renewcommand{\theequation}{\arabic{section}.\arabic{equation}}
\newcommand{\Section}[1]{\section{#1}
\setcounter{equation}{0}}
\newcommand{\proofover}{\hfill\vrule height8pt width8pt depth 0pt}

\newcommand{\secref}[1]{Section~\ref{#1}}
\newcommand{\appref}[1]{Appendix~\ref{#1}}
\newcommand{\eqnref}[1]{(\ref{#1})}
\newcommand{\figref}[1]{Figure~\ref{#1}}
\newcommand{\thmref}[1]{Theorem~\ref{#1}}
\newcommand{\defref}[1]{Definition~\ref{#1}}

\newcommand{\corref}[1]{Corollary~\ref{#1}}

\title{One--bit Distributed Sensing and Coding for \\Field Estimation in
Sensor Networks
}

\author{Ye Wang, Prakash Ishwar, and Venkatesh
Saligrama$^\dagger$
\thanks{$^\dagger$ Y.~Wang, P.~Ishwar, and V.~Saligrama are with the
        Department of Electrical and Computer Engineering,
        Boston University, Boston, MA 02215. Email:
        {\tt\small \{yw,pi,srv\}@bu.edu}.}
}

\begin{document}

\maketitle
\thispagestyle{plain}
\pagestyle{plain}

\begin{abstract}
This paper formulates and studies a general distributed field
reconstruction problem using a dense network of noisy one--bit
randomized scalar quantizers in the presence of additive observation
noise of unknown distribution.  A constructive quantization, coding,
and field reconstruction scheme is developed and an upper--bound to
the associated mean squared error (MSE) at any point and any snapshot
is derived in terms of the local spatio--temporal smoothness
properties of the underlying field. It is shown that when the noise,
sensor placement pattern, and the sensor schedule satisfy certain weak
technical requirements, it is possible to drive the MSE to zero with
increasing sensor density at points of field continuity while ensuring
that the per--sensor bitrate and sensing--related network overhead
rate simultaneously go to zero.  The proposed scheme achieves the
order--optimal MSE versus sensor density scaling behavior for the
class of spatially constant spatio--temporal fields.
\end{abstract}

\Section{\label{sec:intro}Introduction and Overview} We study the
problem of reconstructing, at a data fusion center, a temporal
sequence of spatial data fields, in a bounded geographical region of
interest, from finite bit--rate messages generated by a dense
noncooperative network of sensors. The data--gathering sensor network
is made up of noisy low--resolution sensors at known locations that
are statistically identical (exchangeable) with respect to the sensing
operation.  The exchangeability
assumption reflects the property of an unsorted collection of
inexpensive mass--produced sensors that behave in a statistically
identical fashion. We view each data field as an unknown deterministic
function over the geographical space of interest and make only the
weak assumption that they have a known bounded maximum dynamic
range. The sensor observations are corrupted by bounded, zero--mean,
additive noise which is independent across sensors with arbitrary
dependencies across field snapshots. This {\em noise has an arbitrary,
unknown distribution} but a known maximum dynamic range. The sensors
are equipped with binary analog--to--digital converters (ADCs) in the
form of comparators with random thresholds which are uniformly
distributed over the (known) sensor dynamic range. These thresholds
are assumed to be independent across sensors with arbitrary
dependencies across snapshots. These modeling assumptions partially
account for certain real--world scenarios that include (i) the
unavailability of good initial statistical models for data fields in
yet to be well studied natural phenomena, (ii) unknown additive
sensing/observation noise sources, (iii) additive model perturbation
errors, (iv) substantial variation of preset comparator thresholds
accompanying the mass--manufacture of low--precision sensors, (v)
significant temperature fluctuations across snapshots affecting
hardware characteristics, and (vi) the use of intentional dither
signals for randomized scalar quantization.

Building upon prior results in
\cite{MasryC-IT1981-BPCNT,Masry-IT1981-RASFS}, and
\cite{Luo-IT2005-UDEBCSN}, we develop a simple coding and field
reconstruction scheme based on one--bit scalar quantized samples of
noisy observations. We characterize the associated scaling behavior of
the MSE of field reconstruction with sensor density in terms of the
local and global moduli of continuity of the underlying sequence of
fields. This MSE characterization is for fixed, positive, and equal
sensor coding rates (bits per sensor per snapshot).  These achievable
results reveal that for bounded, zero--mean, additive observation
noise of unknown distribution, the MSE at every point of continuity of
every field snapshot can be made to go to zero as sensor density
increases while simultaneously sending the per--sensor bitrate and any
sensing--related network rate overheads (e.g., sensor addresses) to
zero.  This is possible if the sensor placement and sampling schedule
satisfy a certain uniformity property. This property ensures that the
field estimate at any given spatial location is formed using the
observations from increasingly many sensors that are located within a
vanishingly smaller neighborhood of the location.

The MSE results of this work pertain to uniform pointwise convergence
to zero, that is, for every spatial location of every field, unlike
results pertaining to spatially and temporally averaged MSE which are
more commonly encountered.  The rate of decay of field reconstruction
MSE at a given location is related to the local modulus of continuity
of the field at the given location and time.  Specializing these
results to the case of spatially constant fields yields an achievable
MSE decay rate of $O(1/N)$ where $N$ is the sensor network
size.\footnote{Landau's asymptotic notation: $f(N) = O(g(N))
\Leftrightarrow \lim\sup_{N\rightarrow \infty}|f(N)/g(N)| < \infty$;
$f(N) = \Omega(g(N)) \Leftrightarrow g(N) = O(f(N))$; $f(N) =
\Theta(g(N)) \Leftrightarrow f(N) = O(g(N))\ \text{and}\ g(N) =
O(f(N))$.}  A Cram\'{e}r--Rao lower--bound on the MSE for parameter
estimation establishes that the $O(1/N)$ MSE scaling behavior is
order--optimal in a minimax sense. Since in our problem formulation,
the per--sensor bitrate is held fixed and equal across sensors, in a
scaling sense, the MSE decreases inversely with the total network
rate.

Previous estimation--theoretic studies of one--bit distributed field
reconstruction have focused on reconstructing a single field snapshot
and have either (i) assumed zero observation noise
\cite{MasryC-IT1981-BPCNT,Masry-IT1981-RASFS}, or (ii) assumed a
spatially constant field (equivalent to scalar parameter estimation)
with a one--bit communication as opposed to a one--bit sensing
constraint \cite{Luo-IT2005-UDEBCSN}. The system proposed in this work
integrates the desirable field sensing and reconstruction properties
of these apparently different one--bit field estimation schemes and
establishes the statistical and performance equivalence of these
approaches. An important hardware implication of this paper is that
noisy op--amps (noisy threshold comparators) are adequate for
high--resolution distributed field reconstruction. This should be be
contrasted with the framework in \cite{Luo-IT2005-UDEBCSN} which
implicitly requires sensors to have the ability to quantize their
observations to an arbitrarily high bit resolution. A side
contribution of this paper is the holistic treatment of the general
distributed field--reconstruction problem in terms of (i) the field
characteristics, (ii) sensor placement characteristics, (iii) sensor
observation, quantization, and coding constraints with associated
sensing hardware implications, (iv) transmission and sensing--related
network overhead rates, and (v) reconstruction and performance
criteria. We have attempted to explicitly indicate and keep track of
what information is known, available, and used where and what is not.

The randomized scalar quantization model for the sensor comparators
not only captures poor sensing capabilities but is also an enabling
factor in the high--fidelity reconstruction of signals from quantized
noisy observations. As shown in \cite{MarcoDLN-IPSN2003-MTCDWSN} in an
information--theoretic setting, and alluded to in
\cite{Masry-IT1981-RASFS}, the use of {\em identical} deterministic
scalar--quantization (SQ) in all sensors will result in the MSE
performance being fundamentally limited by the precision of SQ, {\em
irrespective of increasing sensor density}, even in the absence of
sensor observation noise.\footnote{The problem will persist even for
identical block vector--quantization (VQ) with identical binning
(hashing) operations.} However, our results further clarify that
having ``diversity'' in the scalar quantizers, achieved, for example,
through the means of an intentional random dither, noisy threshold, or
other mechanisms, can achieve MSE performance that tends to zero as
the density of sensors goes to infinity (\secref{sec:MSEanalysis},
Implications).  Randomization enables high--precision signal
reconstruction because zero--mean positive and negative fluctuations
around a signal value can be reliably ``averaged out'' when there are
enough independent noisy observations of the signal value.  This
observation is also corroborated by the findings reported in the
following related studies \cite{Masry-IT1981-RASFS,
MasryC-IT1981-BPCNT,
Luo-IT2005-UDEBCSN,zorands2000,zorandli2002,IshwarKR-2003-DSDSNBCP,KumarIR-2004-DSSNBLF}.

The results of this work are also aligned with the
information--theoretic, total network rate versus MSE scaling results
for the CEO problem which was first introduced in
\cite{BergerZV-IT1996-TCP} and thereafter studied extensively in the
information theory literature (see
\cite{ViswanathanB-IT1997-QGCP,PrabhakaranTR-ISIT04-RQGCP} and
references therein). However, it should be noted that
information--theoretic rate--distortion studies of this and related
distributed field reconstruction (multiterminal source coding)
problems typically consider stationary ergodic stochastic fields with
complete knowledge of the field and observation--noise statistics,
block--VQ and binning operations, and time and space--averaged (as
opposed to worst--case) expected distortion criteria. In VQ, sensors
are allowed to collect long blocks of real--valued field samples (of
infinite resolution) from multiple field snapshots before a discrete,
finite bit--rate VQ operation.  The fields are often assumed to be
spatially constant and independent and identically distributed (iid)
across time (frequently Gaussian) and the observation noise is often
assumed to be additive with a known distribution (frequently Gaussian)
as in the CEO problem. It should also be noted that the MSE scaling
results for the CEO problem in \cite{ViswanathanB-IT1997-QGCP} are
with respect to the total network rate where the number of agents (or
sensors) has already been sent to infinity while maintaining the total
network rate a finite value. Recent information--theoretic results for
stationary fields under zero observation noise have been developed in
\cite{KashyapLXL-2005-DSCDSN,NeuhoffP-2006-UPRIDLSC}.  There is also a
large body of work on centralized oversampled A--D conversion, e.g.,
see \cite{Cvetkovic-IT2003-RPREUANQ} and references therein.  Our work
does not explicitly address physical--layer network data transport
issues. In particular, we do not consider joint source--channel coding
strategies (however see remark before
Section~\ref{sec:deploymentStrats}). For certain types of joint
source--channel coding aspects of this and related problems, we refer
the reader to the following references
\cite{GastparV-2003-SCCSN,GastparRV-2003-TCNCLSCCR,NowakMW-2004-EIFWSN,LiuES-2005-OPFESN,BajwaSN-2005-MSCCFEWSN,LiuU-2006-ODPTGSN}.
Networking issues such as sensor scheduling, quality of service, and
energy efficiency may be found in \cite{ZhaoST-2006-IBSPNSN} and
references therein.

The rest of this paper is organized as follows. The main problem
description with all the associated technical modeling assumptions is
presented in \secref{sec:problemsetup}. The main technical results of
this paper are then crisply summarized and their implications are
discussed in \secref{sec:results}.  \secref{sec:ourscheme} describes
the proposed constructive distributed coding and field reconstruction
scheme and the analysis of MSE performance which leads to the
technical results of \secref{sec:results}. For completeness, in
Section~\ref{sec:deploymentStrats} we also briefly discuss sensor
deployment issues but this is not the focus of this work. In
\secref{sec:prevresults}, we discuss the close connections between the
work in \cite{Masry-IT1981-RASFS}, \cite{Luo-IT2005-UDEBCSN}, and the
present work, and establish the fundamental statistical and
performance equivalence of the core techniques in these studies. We
also discuss how the scenario of arbitrary unbounded noise and
threshold distributions can be accommodated when the statistics are
known.  We conclude in \secref{sec:conclusions} by summarizing the
main findings of this work. The proofs of two main results are
presented in the appendices.

\Section{\label{sec:problemsetup}Distributed Field Reconstruction
Sensor--network (DFRS) Setup}

\begin{figure*}
\centering
\includegraphics[width=7.0in]{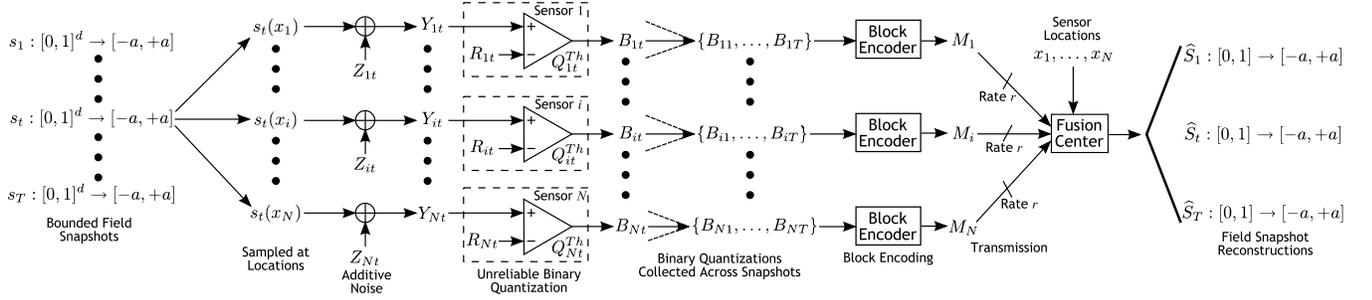}
\caption{\label{fig:probSetup} {\bf Block diagram of a distributed
field reconstruction sensor--network using randomized $1$--bit SQ
with block--coding.}  Sensor $i$ quantizes its noisy observations,
$Y_{i1}, \ldots, Y_{iT}$, to the binary values $B_{i1}, \ldots,
B_{iT}$. The sensor then generates the message $M_i \in \{1, \ldots,
2^{rT}\}$ based on these quantized values. These messages $\{M_i\}$
are then relayed to the fusion center where the field estimates
$\widehat{S}_t$ are produced.}
\end{figure*}

\subsection{\label{sec:fieldModel}Field Model}

We consider a sequence of $T$ discrete--time snapshots of a
spatio--temporal field.\footnote{If the spatio--temporal field is
temporally bandlimited then the field values at intermediate time
points can be interpolated from the estimates at discrete time
snapshots if the temporal sampling rate is (strictly) higher than
the temporal Nyquist rate of the field. The associated MSE will be
no larger than the maximum MSE of the estimates across the
discrete--time snapshots times a proportionality constant.} Each
snapshot is modeled as a continuous\footnote{More generally, our
results can be extended to arbitrary, amplitude--bounded, measurable
functions. For such functions the pointwise MSE bounds given in
\secref{sec:MSEanalysis} still hold. The estimates at the points of
continuity will have MSE tending to 0 as the network size scales.
However, the points of discontinuity may have a finite, but
non--zero MSE floor.} bounded function,
\[
s_t:G \rightarrow \rR:\ \forall x \in G,\ \forall t \in
\{1,\ldots,T\},\ |s_t(x)| \leq a < +\infty,
\]
where $G \subseteq \rR^d$ is a known geographical region of interest
in $d$--dimensional real space and $a$ is a known bound on the
maximum field dynamic range. Although the results of this paper hold
for any $G$ which is bounded and is the closure of its nonempty
interior, for simplicity and clarity of exposition, we will assume
$G = [0,1]^d$, the $d$--dimensional unit--hypercube, in the sequel.
Distances are measured with respect to a norm\footnote{For
asymptotic results in which distance $\longrightarrow 0$, any norm
on $\rR^d$ would suffice since all norms on any finite--dimensional
Banach space are equivalent \cite[Theorem~23.6,
p.~177]{AliprantisB-AP90-PRA}.} $\|\cdot\|$, which for this work
will be assumed to be the Euclidean $2$--norm. Since the fields are
continuous functions on the compact set $G$, they are in fact
uniformly continuous on $G$ \cite{AliprantisB-AP90-PRA}.

Results on the fidelity of the field reconstruction will be
described in terms of the local and global moduli of continuity
associated with the field:

\begin{definition}\label{def:localMod}\emph{(Local modulus of
continuity)} The local modulus of continuity $\omega_t:[0,\infty)
\times G \rightarrow [0,\infty)$ of the function $s_t(x)$ at the
point $x \in G$ is defined as
\[
\omega_t(\delta,x) \triangleq \sup_{\{x^\prime \in G:\|x-x^\prime\|
\leq \delta\}} |s_t(x) - s_t(x^\prime)|.
\]
Note that for all $x \in G$, $\omega_t(\delta,x)$ is a nondecreasing
function of $\delta$ and that it $\longrightarrow 0$ as $\delta
\longrightarrow 0$ since $s_t(x)$ is continuous at each point $x$ in
$G$.
\end{definition}

\begin{definition}\label{def:globalMod} \emph{(Global modulus of continuity)}
The global modulus of continuity $\widetilde{\omega}_t: [0,\infty)
\rightarrow [0,\infty)$ of the function $s_t(x)$ is defined as
\[
\widetilde{\omega}_t(\delta) \triangleq \sup_{x \in G}
\omega_t(\delta,x).
\]
Again note that $\widetilde{\omega}_t(\delta)$ is a nondecreasing
function of $\delta$ and that it $\longrightarrow 0$ as $\delta
\longrightarrow 0$ since $s_t(x)$ is uniformly continuous over $G$.
\end{definition}

The global and local moduli of continuity of a spatial field
respectively reflect the degree of global and local spatial
smoothness of the field with smaller values, for a fixed value of
$\delta$, corresponding to greater smoothness. For example, for a
spatially constant field, that is, for all $x\in G$, $s_t(x) = s_t
\text{ (a constant)}$, we have $\widetilde{\omega}_t(\delta) = 0$
for all $\delta \geq 0$. For $d=1$ and fields with a uniformly
bounded derivative, that is, for all $x \in G$, $\sup_{x\in
G}|d(s_t(x))/dx| = \Delta < +\infty$, $\widetilde{\omega}_t(\delta)
\leq \Delta \cdot \delta$. More generally, for a Lipschitz--$\gamma$
spatial function (see \cite{MasryC-IT1981-BPCNT}) $s_t(x)$, we have
$\widetilde{\omega}_t(\delta) \propto \delta^\gamma$. Closed--form
analytical expressions of moduli of continuity may not be available
for arbitrary fields but bounds often are. Sometimes bounds that are
tight in the limit as $\delta \longrightarrow 0$ are also available.
From Definitions~\ref{def:localMod}, \ref{def:globalMod}, and the
boundedness of the field dynamic range, it also follows that for all
$\delta \geq 0$, for all $x \in G$, and for all $t \in
\{1,\ldots,T\}$, we have
\begin{eqnarray*}
0 \leq \omega_t(\delta,x) \leq \widetilde{\omega}_t(\delta) \leq 2a
< +\infty.
\end{eqnarray*}

\subsection{\label{sec:sensePlace}Sensor Placement}

We assume that we have a dense, noncooperative network of $N$ sensors
distributed uniformly over a hypercube partitioning of $G =
[0,1]^d$. The space $G = [0,1]^d$ is uniformly partitioned into $L =
l^d$ (where $l$ is an integer) disjoint, hypercube supercells of
side--length $(1/l)$. Each supercell is then further uniformly
partitioned into $M = m^d$ (where $m$ is an integer) hypercube
subcells of side--length $(1/(lm))$, giving a total of $LM$
subcells. In our distributed field coding and reconstruction scheme,
described in \secref{sec:ourscheme}, the field estimate for each
snapshot is constant over each supercell and is formed by averaging
the measurements from a partial set of the sensors determined by the
subcells. This field reconstruction scheme requires knowledge of the
sensor locations only up to supercell (not subcell)
membership. Therefore, it has some natural robustness against sensor
location uncertainty or error. The significance of the super and
subcells will become clear in the sequel (Sections~\ref{sec:results}
and \ref{sec:ourscheme}).

We assume that the sensor deployment mechanism is able to uniformly
distribute the sensors over the subcells. We define this uniform
sensor deployment condition with:

\begin{definition}\label{def:UnifPlacement} \emph{(Uniform sensor deployment)}
We say that a sensor deployment method is uniform if exactly $n
\triangleq (N/(LM))$ sensors are located in each subcell.
\end{definition}

\defref{def:UnifPlacement} describes ideal sensor deployment
conditions and can be achieved by locating the sensors over a uniform
grid. However, precise control of sensor locations may not be possible
in practice.
Since we are not primarily concerned about the details of deployment,
we defer discussion of such issues to \secref{sec:deploymentStrats},
where we introduce a stochastic deployment model in order to capture
the uncertainty of realistic deployment mechanisms. In
\secref{sec:deploymentStrats}, we show that this deployment method
satisfies a relaxed version of \defref{def:UnifPlacement}, the
asymptotic nearly uniform deployment condition given by
\defref{def:NearUnifPlacement}, which does not significantly change
the estimator performance.

\begin{figure}
\centering
\includegraphics[width=3.0in]{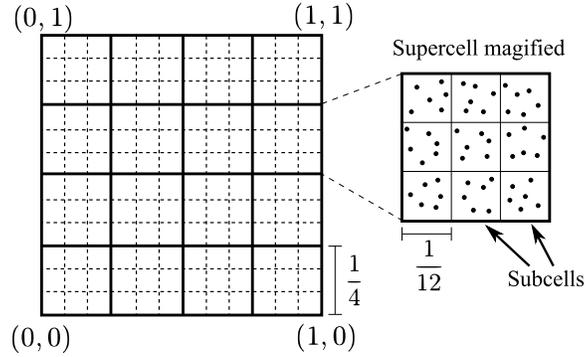}
\caption{{\bf Example uniform sensor deployment and cell hierarchy
over $[0,1]^2$ ($d = 2$).} Here, $N = 864$ sensors are deployed over
$L = 4^2$ supercells of side--length $(1/4)$ and $M = 3^2$ subcells
per supercell of side--length $(1/(3 \cdot 4))$, resulting in $6$
sensors per subcell.} \label{fig:deployment}
\end{figure}

For clarity of presentation, we will assume that the deployment scheme
being used satisfies the uniform sensor deployment condition given in
\defref{def:UnifPlacement}. We also assume that each sensor is aware
of which subcell it is in. \figref{fig:deployment} illustrates the
cell hierarchy and an example sensor deployment for the $d = 2$
dimensional case.

\subsection{Sensor Observation and Coding Models}

\subsubsection{\label{sec:obsModel}Sensor Observation Noise}

The sensor observations are corrupted by bounded, zero--mean
additive noise which is independent across sensors, but can be
arbitrarily correlated across field snapshots\footnote{The
measurement snapshot timers of all the participating sensors are
assumed to be synchronized.}. Let $Z_{it}$ denote the noise
affecting the observation of the $t^{\mathrm{th}}$ snapshot by the
$i^{\mathrm{th}}$ sensor, and define the $\mathbf{Z} \triangleq
\{Z_{it}\}_{i=1,t=1}^{N,T}$ (the collection of all of the noise
random variables) and $\mathbf{Z}_i \triangleq \{Z_{it}\}_{t=1}^{T}$
(the collection of all of the noise random variables for a given
sensor $i$). The noise $\mathbf{Z}$ has an unknown joint cumulative
distribution function (cdf) $F_{\mathbf{Z}}(\mathbf{z})$ that can be
arbitrary within the zero--mean, boundedness and independence
constraints already stated. The maximum dynamic range of the noise
$b \in [0,+\infty)$ is known. The noisy observation of field
snapshot $t \in \{1, \ldots, T \}$ made by sensor $i \in \{1,
\ldots, N\}$ is given by
\[
Y_{it} = s_t(x_i) + Z_{it},
\]
where $x_i$ is the location of the $i^\mathrm{th}$ sensor and
$\mathbf{Z} \sim \mbox{cdf } F_{\mathbf{Z}}(\mathbf{z})$. We use
$\mathcal{F}$ to denote the set of all joint cdfs that are
factorizable into $N$ zero--mean joint cdfs on $\rR^T$ with support
within $[-b,+b]^T$, that is, $F_{\mathbf{Z}}(\mathbf{z}) =
\prod_{i=1}^N F_{\mathbf{Z_i}}(\mathbf{z_i})$ where
$F_{\mathbf{Z_i}}(\mathbf{z_i})$ is a zero--mean joint cdf
(corresponding to the noise random variables for sensor $i$) with
support within $[-b,+b]^T$. Note that $\mathcal{F}$ captures the
feasible set of joint noise cdfs for the bounded--amplitude,
zero--mean, and independence assumptions. Also note that $|Y_{it}|
\leq |s_t(x_i)| + |Z_{it}| \leq c \triangleq (a+b)$.

\subsubsection{\label{sec:1bitSQ}Randomized $1$--bit SQ with Block Coding}

Due to severe precision and reliability limitations, each sensor $i
\in \{1, \ldots, N\}$, has access to only to a vector of unreliable
binary quantized samples $\mathbf{B}_i \triangleq (B_{i1}, \ldots,
B_{iT})$ for processing and coding and not direct access to the
real--valued noisy observations $Y_{i1}, \ldots, Y_{iT}$. The
quantized binary sample $B_{it}$ is generated from the corresponding
noisy observation $Y_{it}$ through a randomized mapping $Q_{it}:
[-c,c] \rightarrow \{0,1\}$: for each $i \in \{1, \ldots, N\}$ and
each $t \in \{1, \ldots, T \}$,
\[
B_{it} = Q_{it}(Y_{it}),
\]
where we assume that the mappings $Q_{it}$ are independent across
sensors $i$, but can be arbitrarily correlated across snapshots $t$.
We denote the conditional marginal statistics of the quantized
samples by $p_{B_{it}|Y_{it}}(y) \triangleq \pP(B_{it} = 1|Y_{it} =
y)$. We are specifically interested in cases where
$p_{B_{it}|Y_{it}}(y)$ is an affine function of $y$ since it allows
estimates of the fields to be made from the $B_{it}$'s without
knowledge of the noise distribution (see \appref{app:MSEperfProof}).
Specifically we consider the conditional distribution
\[
p_{B_{it}|Y_{it}}(y) = \left(\frac{y + c}{2c}\right).
\]

This conditional distribution can be achieved by a quantization
method which is based on comparing the noisy observation with a
random uniformly distributed threshold given by
\begin{equation}\label{eqn:ourQFunc}
B_{it} = Q_{it}^{Th}(Y_{it}) \triangleq \mathbf{1}(Y_{it} > R_{it}),
\end{equation}
where the $R_{it}$'s are $\mathrm{Unif}[-c,c]$ random thresholds
which are independent across sensors $i$, but arbitrarily correlated
across snapshots $t$, and $\mathbf{1}(\cdot)$ denotes the indicator
function:
\[
\mathbf{1}(Y_{it} > R_{it}) =
\begin{cases} 1 & \mbox{if } Y_{it} > R_{it}, \\
0 & \mbox{otherwise}.
\end{cases}
\]

This uniform random--threshold $1$--bit SQ model partially accounts
for some practical scenarios that include (i) comparators with a
floating threshold voltage, (ii) substantial variation of preset
comparator thresholds accompanying the mass--manufacture of
low--precision sensors, (iii) significant environmental fluctuations
that affect the precision of the comparator hardware, or generally
(iv) unreliable comparators with considerable sensing noise and
jitter. An alternative justification is that the random thresholds
are intentionally inserted as a random dither. Scenario (i) can be
accommodated by independence across snapshots, scenario (ii) can be
accommodated by complete correlation (fixed) across snapshots, and
scenarios (iii) and (iv) can be accommodated by arbitrary
correlation across snapshots.

\begin{figure}[!htb]
\centering
\includegraphics[width=2.5in]{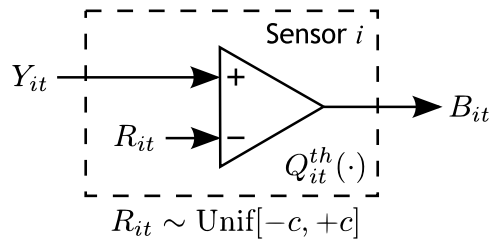}
\caption{\label{fig:QthModel}{\bf Quantizer hardware example.} The
sensing model described by the $Q_{it}^{Th}(\cdot)$ function in
\eqnref{eqn:ourQFunc} can be implemented by a comparator with a
uniformly distributed threshold. These thresholds are independent
across sensors, but arbitrarily correlated across snapshots,
allowing many scenarios to be accommodated.}
\end{figure}

Each sensor $i$ utilizes a block encoder to ``compress'' its vector
of $T$ quantized samples $\mathbf{B}_i$ to a message $M_i \in \{1,
2, \ldots, 2^{rT}\}$ before transmitting to the fusion center.  The
block encoder and message for sensor $i$ are given by
\[
f_i:\{0,1\}^T \rightarrow \{1, 2, \ldots, 2^{rT}\}, \quad M_i =
f_i(B_{i1}, \ldots, B_{iT}),
\]
where $r$ is the coding rate in bits per sensor per snapshot. For $r
\geq 1$ compression is trivial since $\mathbf{B}_i$ can assume no
more than $2^T$ distinct values which can be indexed using $T$~bits.

\subsection{\label{sec:transModel}Transmission and Field Reconstruction}

In this work, a data fusion center is any point of data aggregation
and/or processing in the sensor network and can be real or virtual.
For instance, sensors can be dynamically organized into clusters with
different sensors assuming the role of a fusion center at different
times \cite{ChouPR-Asilomar2002-TECDSN}.  To conform with the existing
base of digital communication architectures, our problem setup
abstracts the underlying transmission network of sensors effectively
as a network of bit pipes. These bit pipes are capable of reliably
delivering these $N$ messages (the payloads) and the network addresses
of the message origination nodes (the headers) to the fusion
center. This enables the fusion center to correctly associate the
spatial location information with the corresponding sensor
field--measurement information for reliable field reconstruction. In
practice, sensor data can be moved to the fusion center through a
variety of physical--layer transport mechanisms, example, a stationary
base--station with directional antenna arrays, a mobile data
collector, and passive sensor querying mechanisms involving, for
instance, laser--beams and modulating mirrors
\cite{KahnKP-MOBICOM99-NCCMNSD}.

Separating the distributed field reconstruction problem into efficient
data acquisition and efficient data transport parts through a
finite--rate reliable bit--pipe abstraction may be suboptimal
\cite[p.~449]{CoverJ-1991-EoIT}, \cite{GastparV-2003-SCCSN,
GastparRV-2003-TCNCLSCCR}. For instance, in some scenarios multihop
communication is not needed and the characteristics of the field, the
communication channel, and the distortion--metric are ``matched'' to
one another. In such a scenario, uncoded ``analog'' transmission can
offer huge performance gains if the synchronization of sensor
transmissions can be orchestrated at the physical layer to achieve
beamforming gains and the network channel state information is
available to the transmitting sensors \cite{GastparV-2003-SCCSN}.
Certain aspects of this analog transmission can be incorporated within
our field reconstruction framework and is briefly discussed in the
remark just before Section~\ref{sec:deploymentStrats}.

For our reconstruction scheme, described in \secref{sec:ourscheme},
the fusion center only needs to be able to spatially localize the
origin of each message to within the supercell resolution.
This can be achieved by having each sensor append a $\log(LM)$ bits
long label to its message. This results in a total sensor--location
rate--overhead of $r_{ohd} = (N/T)\log(LM)$ bits per snapshot on the
network information transport costs. This overhead will be negligible
if $T \gg N\log(LM)$. If the underlying sequence of fields are
spatially constant, then, the sensor location information is not
needed at the fusion center (see \corref{cor:constFieldCase} and
\secref{sec:ourscheme}).

The fusion center forms the estimates of the $T$ fields based on the
sensor messages using the reconstruction functions
\[
g_t: G \times \{1, 2, \ldots, 2^{rT}\}^N \rightarrow [-a,a], \
\forall t \in \{1, \ldots, T\}.
\]
The estimate of field $t$ at point $x \in G$ is denoted by
\[
\widehat{S}_t(x) = g_t(x, M_1, \ldots, M_N).
\]
\begin{definition}\emph{(Rate--$r$ DFRS)}
A rate--$r$ DFRS based on randomized $1$--bit SQ with block coding
is defined by the set of rate--$r$ encoder functions
$\{f_i(\cdot)\}_{i=1}^N$ and the set of reconstruction functions
$\{g_t(\cdot)\}_{t=1}^T$.
\end{definition}

Figure~\ref{fig:probSetup} depicts a rate--$r$ DFRS using randomized
$1$--bit SQ with block coding.

\subsubsection{\label{sec:perfCriteria}Performance Criterion}

\begin{definition} \emph{(Pointwise MSE)}
The pointwise MSE of the estimate of field $t$ at location $x \in
G$, for a given rate--$r$ DFRS and a specific noise joint cdf
$F_\mathbf{Z}(\mathbf{z}) \in \mathcal{F}$, is given by
\[
D_t(x;F_\mathbf{Z}) = \eE[(\widehat{S}_t(x) - s_t(x))^2].
\]
\end{definition}

Since we are interested in schemes that will work for {\em any}
noise cdf in $\mathcal{F}$, we consider the worst--case
$D_t(x;F_\mathbf{Z})$ over all possible $F_\mathbf{Z} \in
\mathcal{F}$. We also consider the maximization over all fields and
all locations in $G$ since we want to reconstruct every point of
every field with high fidelity.
\begin{definition}\label{def:worsecaseMSE} \emph{(Worst--case MSE)}
The worst--case MSE $D$ is given by
\[
D = \max_{t \in \{1, \ldots, T\}} \sup_{x \in G} \sup_{F_\mathbf{Z}
\in \mathcal{F}} D_t(x;F_\mathbf{Z}).
\]
\end{definition}

Our objective is to understand the scaling behavior of MSE with $N$,
$T$, and $r$. The next section summarizes our partial results in
this direction.

\Section{\label{sec:results}Main Results}

\subsection{\label{sec:MSEanalysis}Achievable MSE Performance}

Our first result gives an upper bound on the MSE achievable through
a constructive DFRS based on randomized $1$--bit SQ with block
coding for rate $r = 1/M$, where $M$ is the number of subcells per
supercell. The actual scheme will be described in
\secref{sec:ourscheme}. The MSE analysis appears within the proof of
the theorem detailed in \appref{app:MSEperfProof}. This achievable
MSE upper bound can be made to decrease to zero as sensor--density
goes to infinity (see \eqnref{eqn:LNScaling}) without knowledge of
the local or global smoothness properties of the sequence of fields.
Furthermore, this scheme is universal in the sense that it does not
assume knowledge of $F_\mathbf{Z}(\mathbf{z})$ beyond membership to
$\mathcal{F}$.

\begin{theorem}\label{thm:MSEperf} \emph{(Achievable MSE performance:
Randomized $1$--bit SQ and $r = 1/M$)} There exists a rate--$r = 1/M$
DFRS based on randomized $1$--bit SQ with block coding (e.g., the
scheme of \secref{sec:ourscheme}) such that for all $x \in G$, $t \in
\{1, \ldots, T\}$, and $F_\mathbf{Z}(\mathbf{z}) \in \mathcal{F}$,
\begin{eqnarray*}
D_t(x;F_\mathbf{Z}) &\leq&
\omega_t^2\left(\frac{\sqrt{d}}{\sqrt[d]{L}} ,x\right) +
\left(\frac{LMc^2}{N}\right) \\
&\leq&
\widetilde{\omega}_t^2\left(\frac{\sqrt{d}}{\sqrt[d]{L}}\right) +
\left(\frac{LMc^2}{N}\right).
\end{eqnarray*}
\end{theorem}

\begin{proof}
See Section~\ref{sec:ourscheme} and \appref{app:MSEperfProof}.
\end{proof}

Note that Theorem~\ref{thm:MSEperf} holds for arbitrary fields. The
modulus of continuity terms in the local (first) and global (second)
upper bounds of Theorem~\ref{thm:MSEperf} are due to the bias of the
field estimates and the $\left(\frac{LMc^2}{N}\right)$ term is due
to the variance of the field estimates (see \eqnref{eqn:1bitreconst}
in \secref{sec:ourscheme}). From \thmref{thm:MSEperf} and the
properties of moduli of continuity (see \secref{sec:fieldModel}), it
follows that for the coding and reconstruction scheme of
\secref{sec:ourscheme}, as $N \longrightarrow \infty$, the estimate
$\widehat{S}_t(x)$ uniformly converges, in a mean square sense, to
$s_t(x)$ for all $x \in G$, provided that
\begin{equation} \label{eqn:LNScaling}
\mbox{(i) }\left(\frac{N}{L}\right) \longrightarrow \infty, \mbox{
and (ii) } L \longrightarrow \infty.
\end{equation}
It also follows that the worst--case MSE scaling behavior (see
Definition~\ref{def:worsecaseMSE}) is bounded by
\begin{equation} \label{eqn:MSE-WC-Result}
D \leq \max_{t \in \{1, \ldots, T\}} \left\{
\widetilde{\omega}_t^2\left(\frac{\sqrt{d}}{\sqrt[d]{L}}\right) +
\left(\frac{LMc^2}{N}\right)\right\}
\end{equation}
and that $D \longrightarrow 0$ as $N$ and $L$ scale as in
\eqnref{eqn:LNScaling}.

\noindent {\bf Implications:} These results allow us to make the per
sensor per snapshot bit rate $r$, worst--case MSE $D$, and sensor
message ID overheads (given by $(N/T) \log(LM)$ bits) simultaneously
smaller than any arbitrarily small desired values $r^*, D^*, \epsilon
> 0$, respectively. First, we can choose a sufficiently large number
of subcells per supercell $M^*$ such that the rate $r = 1/M^* < r^*$.
Then we can choose a sufficiently large number of sensors $N^*$ and
number of supercells $L^*$ such that the bound on $D$ given by
\eqnref{eqn:MSE-WC-Result} is made less than $D^*$. Note that both
$N^*$ and $M^*$ can be further increased while keeping the ratio
$M^*/N^*$ fixed without changing the bound on $D$. This corresponds to
increasing the total number of sensors $N$, decreasing the per sensor
rate $r = 1/M$, but keeping the total network per snapshot rate $Nr =
N/M$ and distortion $D$ fixed. Finally, we can look at a sufficiently
large number of snapshots $T^*$ such that network message overheads
$(N^*/T^*) \log(L^*M^*) < \epsilon$.

In the constructive coding and field reconstruction scheme of
\secref{sec:ourscheme}, the field estimates are piecewise constant
over the supercells. The estimate in each supercell is formed from
only $n = (N/(LM))$ of the $Mn = (N/L)$ quantized observed values
coming from the sensors located in that supercell. Since only
$(1/M)$ of the total available quantized observed values for each
snapshot are used, the transmission rate of $(1/M)$ is achievable by
indexing only the necessary values (see \secref{sec:ourscheme} for
details). As the number of supercells $L$ increases, the piecewise
constant estimate becomes finer and the bias is decreased. Also, as
the number of sensors per supercell is increased, more observations
are used thus decreasing the variance of the estimate.

Since the variance term $\frac{LMc^2}{N}$ in the upper bound of
Theorem~\ref{thm:MSEperf} can decrease no faster than $O(1/N)$, the
decay of the global MSE upper bound, in the proposed constructive
scheme, can be no faster than $O(1/N)$. However, the decay rate of
$\frac{LMc^2}{N}$ is hindered by the fact that $L$ simultaneously
needs to approach infinity for the bias term
$\widetilde{\omega}_t^2\left(\frac{\sqrt{d}}{\sqrt[d]{L}}\right)$ to
decay to $0$. When $\widetilde{\omega}_t(\cdot)$ is not identically
zero, a bias--variance tradeoff exists and the appropriate relative
growth rate for $L$ with $N$ that minimizes the decay rate of the
global MSE upper bound of Theorem~\ref{thm:MSEperf} is determined by
the following condition
\[
\widetilde{\omega}_t^2\left(\frac{\sqrt{d}}{\sqrt[d]{L}}\right) =
\Theta\left(\frac{L}{N}\right).
\]
For certain classes of signals for which the global modulus of
continuity has a closed form, the optimum growth rate can be
explicitly determined. For instance, if $d=1$ and
$\widetilde{\omega}_t(\delta) = \Delta \cdot \delta$ (Lipschitz--$1$
fields), $L_{opt}(N) = \Theta(N^{1/3})$ for which $\mathrm{MSE} =
O(N^{-2/3})$.

\begin{corollary}\label{cor:constFieldCase} \emph{(Achievable MSE
performance: Randomized $1$--bit SQ, $r = 1/M$, and constant
fields)} If for all $x \in G$ and all $t \in \{1, \ldots, T\}$, we
have $s_t(x) = s_t$, or equivalently, for all $\delta \geq 0$ and
all $t \in \{1, \ldots, T\}$, $\widetilde{\omega}_t(\delta) = 0$,
then the result given by \eqnref{eqn:MSE-WC-Result} reduces to
\[
D \leq \left(\frac{Mc^2}{N}\right),
\]
where we can set $L = 1$ to minimize the bound.
\end{corollary}
Only $L = 1$ supercell is needed for an accurate piecewise constant
reconstruction of a constant field. Furthermore, all
snapshot--estimates given by the scheme from \secref{sec:ourscheme}
are unbiased in this case. Also, the spatial locations of sensors are
irrelevant: the MSE behavior is governed purely by the number of
sensors $N$ regardless of how they are distributed over the
subcells. The $N$ sensors must still be uniformly assigned to one of
$M$ groups (for the purpose of transmission coordination to achieve
the compression factor of $1/M$), however these groups need not have
any geographical significance.

The MSE results given by \thmref{thm:MSEperf} show that the field
snapshot estimates converge uniformly in MSE and upper bound the MSE
decay rate. Every point of every estimate, in fact, converges almost
surely to the true value. We also state a central limit theorem
(CLT) result regarding the estimation error.

\begin{theorem}\label{thm:ASConv} \emph{(Almost--sure convergence of field
estimates)} There exists a rate--$r = 1/M$ DFRS based on randomized
$1$--bit SQ with block coding (described in \secref{sec:ourscheme})
such that for all $x \in G$, $t \in \{1, \ldots, T\}$, and
$F_\mathbf{Z}(\mathbf{z}) \in \mathcal{F}$,
\begin{eqnarray*}
\widehat{S}_t(x) \xrightarrow{\mathrm{a.s.}} s_t(x),
\end{eqnarray*}
as $N$ and $L$ scale as given in \eqnref{eqn:LNScaling}.
\end{theorem}
\begin{proof}
See Section~\ref{sec:ourscheme} and \appref{app:ASConvProof}.
\end{proof}

\begin{corollary}\label{cor:errorCLT} \emph{(Central limit theorem for
estimation errors)} For the rate $r = 1/M$ DFRS of
\secref{sec:ourscheme}, the normalized error at point $x \in G$ for
the estimate of field snapshot $t \in \{1, \ldots, T\}$, given by
\begin{eqnarray*}
\frac{\widehat{S}_t(x)-s_t(x)}{\sqrt{\mathrm{var}[\widehat{S}_t(x)-s_t(x)]}},
\end{eqnarray*}
is asymptotically zero--mean, unit--variance, normal as $N$ and $L$
scale as given in \eqnref{eqn:LNScaling}, for any
$F_\mathbf{Z}(\mathbf{z}) \in \mathcal{F}$.
\end{corollary}
\begin{proof}
The proof is similar to and follows directly from the proof of
Theorem~2.4 in \cite{Masry-IT1981-RASFS}.
\end{proof}

\subsection{Order--Optimal Minimax MSE for Constant Fields}
\label{sec:converse}

The minimax reconstruction MSE over the class of constant fields is
given by
\begin{equation*}
\inf_{\{g_t\}_{t=1}^{t=T}} \sup_{F_{\mathbf{Z}} \in \mathcal{F}, s_t
\in \mathcal{S}} D,
\end{equation*}
where the infimum is taken over all possible estimators and the
supremum is taken over all noise distributions and fields from the
class of constant fields which is denoted by $\mathcal{S}$. The
achievable MSE result given by \corref{cor:constFieldCase}
establishes an upper bound on the minimax reconstruction MSE.
\thmref{thm:converse} lower bounds the minimax reconstruction MSE
for any rate $r$ DFRS that produces unbiased estimates for the case
of spatially constant fields.

\begin{theorem}\label{thm:converse}
\emph{(Lower bound on MSE: Unbiased estimators for constant fields)}
For a sequence of spatially constant fields and any DFRS which
produces unbiased field estimates, there exists a joint cdf
$F_{\mathbf{Z}} \in \mathcal{F}$ such that for noise distributed
according to $F_{\mathbf{Z}}$ the MSE is lower bounded by
\[
\eE[(\widehat{S}_t - s_t)^2] \geq \left(\frac{C_t}{N}\right), \quad
\text{for all } t \in \{1, \ldots, T\},
\]
where $C_t$ is finite, non--zero, and does not depend on $N$.
Therefore,
\begin{equation*}
\inf_{\{g_t\}_{t=1}^{t=T}} \sup_{F_{\mathbf{Z}} \in \mathcal{F}, s_t
\in \mathcal{S}} D \geq \max_{t \in \{1, \ldots, T\}}
\left(\frac{C_t}{N}\right).
\end{equation*}
\end{theorem}

\begin{proof}
Since $\{s_t\} \rightarrow \{Y_{it}\} \rightarrow \{B_{it}\}
\rightarrow \{M_i\}$ forms a Markov chain, the estimates based on
the sensor messages $\{M_1, \ldots, M_N\}$ cannot have a lower MSE
than estimates based on the noisy observations $\{Y_{it}\}$. Let
$F_{\mathbf{Z}} \in \mathcal{F}$ be any well--behaved, non--trivial,
joint cdf such that the $Z_{it}$ are iid and the conditional
probabilities of $Y_{it}$ given the fields satisfy the regularity
conditions necessary for the Cram\'{e}r--Rao bound
\cite{Kay-1993-FSSPET} to be applied. By the Cram\'{e}r--Rao bound,
the MSE of each field estimate based on $\{Y_{it}\}$ is lower
bounded by $\frac{C_t}{N}$ where $C_t$ is finite, non--zero, and
depends on $F_{\mathbf{Z}}$, but does not depend on $N$. Note that
the bound also applies to general randomized $1$--bit SQ functions
$Q_{it}(\cdot)$ including those based on uniform random thresholds
$Q_{it}^{Th}(\cdot)$ (see \eqnref{eqn:ourQFunc}).
\end{proof}

Combining the results of \corref{cor:constFieldCase} and
\thmref{thm:converse} establishes that the order--optimal minimax
MSE for spatially constant fields is $\Theta(1/N)$ and that the
scheme of \secref{sec:ourscheme} achieves this order optimal
performance.

\Section{\label{sec:ourscheme}Proposed Constructive Distributed
Coding and Field Reconstruction Scheme}

In this section we present the proposed DFRS scheme that was alluded
to in \secref{sec:results}. In this scheme, sensors create the
quantized binary samples $\{B_{it}\}$ from their observations
$\{Y_{it}\}$ through comparisons with the random thresholds
$\{R_{it}\}$, as described in \eqnref{eqn:ourQFunc} of
\secref{sec:1bitSQ}. The field estimates are piecewise constant over
the supercells, where the estimate formed in each supercell is a
function of only $(N/(LM))$ of the $(N/L)$ quantized observed values
coming from the sensors located in that supercell. This allows
fractional transmission rates of $r = 1/M$ through a simple
time--sharing based compression method. Note that there can be
uncertainty in the sensor locations, within a degree given by the
size of a supercell, at the fusion center, since it is only
necessary for the fusion center to know which supercell each sensor
is located in.

Each sensor $i$, instead of transmitting all of its $T$ bits (the
vector of its binary quantized observations $\mathbf{B}_i = (B_{i1},
\ldots, B_{iT})$), transmits only $rT = T/M$ of them and the remaining
observations are dropped. Or alternatively, the sensor may sleep and
not record the remaining measurements. The two--level hierarchy of
supercells and subcells described in \secref{sec:sensePlace} is used
in order to properly determine which bits sensors should drop or
keep. Within each supercell, each sensor $i$ from subcell $k \in \{1,
\ldots, M\}$ communicates only every $M^\mathrm{th}$ bit (offset by
$k$), that is $\{B_{i,k+Ml} \}_{l = 0}^{l = (T/M)-1}$.  These $rT$
bits can be uniquely represented by the message $M_i \in \{1, \ldots,
2^{rT}\}$ and losslessly communicated to the fusion center. Thus for
snapshot $t \in \{1,\ldots,T\}$, only the bits from senors in the
$[((t-1) \mbox{ mod } M) + 1]^{\mathrm{th}}$ subcell of each supercell
are communicated to the fusion center.  The set of all sensor indices
corresponding to the $n = (N/(LM))$ sensors belonging to the $[((t-1)
\mbox{ mod } M) + 1]^{\mathrm{th}}$ subcell of supercell $j$ will be
denoted by $I(j,t)$.  In other words, this set of indices corresponds
to all those sensors which are located in supercell $j$ and are
responsible for recording and encoding a bit in the $t$-th snapshot.

For notational simplicity, the reconstruction function
$\widehat{S}_t(x) = g_t(x, M_1, \ldots, M_N)$ will be described
directly in terms of the available binary quantized observations
$B_{it}$\footnote{The set of binary quantized observations for
snapshot $t$ which are available at the fusion center is given by
$\{B_{it}\}_{\{i \in \cup_{j=1}^L I(j,t)\}}$} and not the encoded
messages $\{M_i\}$ which are information equivalent. The
reconstruction function $\widehat{S}_t(x)$ is piecewise constant and
is described as follows. The field $s_t(x)$ is reconstructed as a
constant over each supercell $j$. The constant estimate is given by
\begin{eqnarray}
\widehat{S}_{tj} &\triangleq& 2c \left[ \frac{1}{n} \sum_{i \in
I(j,t)} B_{it} \right] - c, \label{eqn:simpleavg}
\end{eqnarray}
which is the simple average (shifted and scaled into $[-c,+c]$) of the
available quantized binary observations of snapshot $t$ from sensors
located in supercell $j$. The overall piecewise--constant estimate for
$s_t(x)$ can be then described as
\begin{eqnarray}
\widehat{S}_t(x) &=& g_t(x, M_1, \ldots, M_N) \nonumber \\
&\triangleq& \sum_{j=1}^L \widehat{S}_{tj} \mathbf{1}(x \in
\mathcal{X}_j), \label{eqn:1bitreconst}
\end{eqnarray}
where $\mathcal{X}_j \subseteq [0,1]^d$ is the set of points within
the $j^{\mathrm{th}}$ hypercube supercell and $\mathbf{1}(x \in
\mathcal{X}_j)$ given by
\[
\mathbf{1}(x \in \mathcal{X}_j) =
\begin{cases} 1 & \mbox{if } x \in \mathcal{X}_j, \\
0 & \mbox{otherwise},
\end{cases}
\]
is the indicator function of the set $\mathcal{X}_j$. Other more
sophisticated reconstruction algorithms are possible. For instance,
instead of the simple average used in (\ref{eqn:simpleavg}), one may
use a weighted average with convex weights, and for the overall
reconstruction in (\ref{eqn:1bitreconst}), one may use a
piecewise--linear or other higher--order interpolation algorithms such
as those based on cubic B--splines (see
\cite{Masry-IT1981-RASFS}). The resulting MSE will be of the same
order. We use the former (simple average, piecewise--constant)
reconstruction because its description and analysis is more compact.
\appref{app:MSEperfProof} proves that the MSE of this constructive
coding and reconstruction scheme is upper bounded by the result
described in Theorem~\ref{thm:MSEperf}.

\noindent{\bf Remark:} As discussed at the beginning of
Section~\ref{sec:transModel}, physical--layer network data transport
issues are not the focus of this work. However, if synchronized analog
beamforming from the sensors within each subcell to the fusion center
can be achieved, then the summation in the reconstruction given by
equation (\ref{eqn:simpleavg}) can be realized directly in the analog
physical layer, by ``adding signals in the air'', using a simple
binary pulse amplitude modulation signaling scheme at each sensor. The
additional estimation error variance due to the receiver amplifier
noise at the fusion center will decrease as $1/n$ by scaling the
sampled received waveform by $1/n$ as in (\ref{eqn:simpleavg}). This
will lead to corresponding achievable power versus distortion
tradeoffs (as opposed to bits versus MSE or sensors versus MSE) which
can be quantified.

\subsection{\label{sec:deploymentStrats}{Sensor Deployment Considerations}}

The conditions given by \defref{def:UnifPlacement} correspond to
exactly $(N/(LM))$ sensors uniformly falling into each subcell with
probability one for all $N$, $L$, and $M$ (ignoring integer
affects). In the regime of perfect sensor placement control (or when
placement error is negligible compared to the width of the cells),
this condition is trivially realized by locating the sensors on a
uniform grid. However, such precise sensor placement control might not
be achievable in practice. In order to address this issue we introduce
a stochastic sensor deployment model, one that captures an extreme
case of (uncontrollable) sensor placement uncertainty where each
sensor is deployed according to a uniform distribution over
$[0,1]^d$. We also relax the uniform sensor placement to an asymptotic
nearly uniform sensor deployment given by
\defref{def:NearUnifPlacement}.



\begin{definition}\label{def:NearUnifPlacement}
\emph{(Asymptotic nearly uniform sensor deployment)} We say that a
sensor deployment method is asymptotically nearly uniform with
parameters $(\gamma,\epsilon,N^*)$ if at least $\gamma n \triangleq
\gamma(N/(LM))$ are located in each subcell with probability at
least $1-\epsilon$ for all $N > N^*$, where $\gamma \in (0,1]$
represents the inverse of the ``over--provisioning'' factor for the
number of sensors needed to be deployed.
\end{definition}

This relaxation does not significantly impact our results since we
are interested in the asymptotic results (as $L$ and $N$ scale as in
\eqnref{eqn:LNScaling}) where the $\gamma$ and $\epsilon$ parameters
of \defref{def:NearUnifPlacement} can be made negligible. Our
stochastic deployment scheme satisfies this asymptotic nearly
uniform condition given in \defref{def:NearUnifPlacement}, and also
almost surely satisfies the uniform deployment condition given by
\defref{def:UnifPlacement} for $N \longrightarrow \infty$.

Consider the scenario where $N$ sensors are deployed iid and uniformly
over $G = [0,1]^d$. The corresponding indices of the subcells (the
total $LM$ subcells can be indexed by an integer from $1$ to $LM$)
that the $N$ sensors fall is denoted by the random sequence
$\mathbf{J} = (J_1, \ldots, J_N)$.
Here, $J_i \sim \mbox{iid } U$, where $U \triangleq (1/(LM), \ldots, 1/(LM))$
is the uniform probability mass function over $LM$ discrete values. We
examine the $N$--type (empirical distribution) $P_{\bf J}^{(N)}$ of
$\mathbf{J}$ in order to examine the level of uniformity in the sensor
deployment. An empirical distribution equal to $U$ corresponds to the
uniform deployment condition of \defref{def:UnifPlacement} being
met. Since the indices are also distributed iid according to $U$, by
the strong law of large numbers, the empirical distribution converges
almost surely to $U$ as $N \longrightarrow \infty$, and thus almost
surely the sensors will be deployed uniformly over the subcells
according to \defref{def:UnifPlacement} as $N \longrightarrow
\infty$.

Also, a result from large deviations theory bounds the probability
that the empirical distribution will not be in a
$\delta$--neighborhood of the uniform distribution. This corresponds
to the event where there exists a subcell that has more than
$\frac{N(1+\delta)}{LM}$ or fewer than $\frac{N(1-\delta)}{LM}$
sensors located within it.
Let $\mathscr{P}^N$ be the set of $N$--dimensional probability
distributions, $U^\delta \triangleq [(1- \delta)/(LM), (1 +
\delta)/(LM)]^N$
be the $\delta$--neighborhood around the uniform probability mass
function $U$, $D(\cdot \| \cdot)$ denote the Kullback--Leibler
distance \cite{CoverJ-1991-EoIT}, and
\[
P^* = \arg \min_{P \in \mathscr{P}^N \setminus U^\delta} D(P \| U)
\]
denote the probability distribution not in $U^\delta$ that is closest
to $U$ in Kullback--Leibler distance.  It should be noted that
$D(P^*\|U) > 0$ for all $\delta > 0$.  Then by Sanov's theorem
\cite[p.~292]{CoverJ-1991-EoIT},
\begin{eqnarray}\label{eqn:sanovBound}
\pP(P_J^N \in \mathscr{P}^N \setminus U^\delta) \leq
(N+1)^{LM}2^{-ND(P^*\|U)} \nonumber \\
= 2^{-N\left(D(P^*\|U)-\frac{LM}{N}\log(N+1)\right)}.
\end{eqnarray}
This inequality bounds the probability that not all subcells have at
least $\frac{N(1-\delta)}{LM}$ sensors within them. This shows that as
long as the number of sensors deployed $N$ grows faster than the
actual number of sensors needed $LM$, then the near uniform deployment
condition will be eventually met. Thus, this determines how many total
sensors $N^* > LM$ need to be deployed in order to satisfy the
asymptotic nearly uniform sampling condition of
\defref{def:NearUnifPlacement} for a given desired $\epsilon$ and for
$\gamma = 1-\delta$.

\Section{\label{sec:prevresults} Discussion of Related One--bit Estimation Problems and Extensions
}

\begin{figure*}
\centering
\includegraphics[width=6.0in]{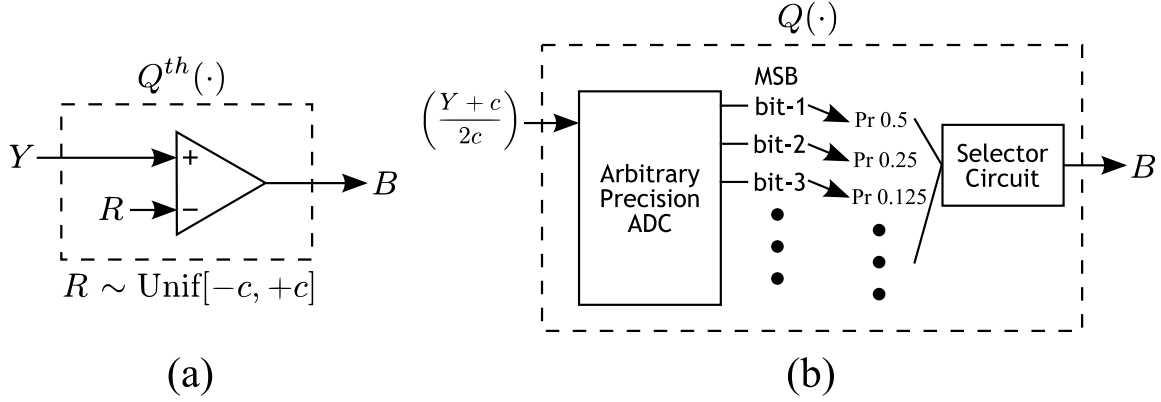}
\caption{\label{fig:compareImp} {\bf The $Q_{it}^{Th}(\cdot)$ function
in \eqnref{eqn:ourQFunc} and the $Q(\cdot)$ function of
\cite{Luo-IT2005-UDEBCSN} suggest markedly different hardware
implementations.} The former naturally suggests (a), where the binary
quantized value is produced by a simple comparison to a random
threshold $X$. The latter suggests (b), where an arbitrarily--precise
ADC circuitry probabilistically selects an arbitrary bit of the
observed value. Interestingly, these two implementations produce
statistically equivalent quantized outputs $B$ given identical inputs
$Y$.}
\end{figure*}

This section discusses the connections between the methods and results
in \cite{Masry-IT1981-RASFS}, \cite{Luo-IT2005-UDEBCSN}, and the
present work. It is shown that the apparently different randomized
$1$--bit field estimation schemes in these studies are in fact
statistically and MSE performance equivalent. We also address how, in
the scenario of known noise statistics, unbounded noise distributions
and arbitrary threshold distributions can be accommodated. The general
framework of the present work integrates the desirable field sensing
and reconstruction properties and insights of the earlier studies and
provides a unified view of the problem that simultaneously considers
unreliable binary quantization, unknown arbitrary noise distributions,
multiple snapshots of a temporally and spatially varying field, and
communication rate issues. Since the work in both
\cite{Masry-IT1981-RASFS} and \cite{Luo-IT2005-UDEBCSN} deal with the
reconstruction of only a single snapshot ($T = 1$), we will drop the
snapshot indices $t$ in our discussion to aid comparison.

\subsection{\label{sec:Masry}One--Bit Randomized--Dithering}


The problem setup of \cite{Masry-IT1981-RASFS} may be viewed as the
reconstruction of a single snapshot ($T = 1$) of a bounded,
one--dimensional field ($d = 1$) from noiseless samples ($Z_i = 0$) at
uniformly spaced (deterministic) sampling locations ($x_i = i/N$).  In
\cite{Masry-IT1981-RASFS} the noiseless observations are binary
quantized using random thresholds $R_i$s that have a known general
distribution which satisfies certain technical conditions described in
\cite[Section~II.A]{Masry-IT1981-RASFS}. These technical conditions
include the uniform distribution (considered in this paper) as a
special case. An important conceptual difference is that in
\cite{Masry-IT1981-RASFS} the sensor quantization noise is viewed as a
randomized dither signal which is intentionally added to the
observations and that the dither cdf is known (it need not be
uniform). The reconstruction explicitly exploits the knowledge of the
dither statistics. Specifically, the noiseless observation $Y_i$, at
sensor $i$, and the corresponding quantized binary sample $B_i$ become
\begin{eqnarray*}
Y_i &=& s(x_i), \\
B_i &=& Q(Y_i) \triangleq \mathrm{sgn}(Y_i + X_i),
\end{eqnarray*}
where $X_i$ is iid dithering noise with a known distribution
$p_X(\cdot)$ which satisfies certain technical assumptions as given
in \cite[Section~II.A]{Masry-IT1981-RASFS}. Note that taking the
sign of the sum of the observation and random dither $X_i$ is
equivalent to comparing with the threshold $-X_i$. Thus the
quantization function $Q(\cdot)$ of \cite{Masry-IT1981-RASFS} is
equivalent\footnote{The sign function maps to $\{-1,+1\}$ whereas a
threshold comparator maps to \{0,1\}. However, the replacement of
the $-1$ symbol with the 0 symbol is unimportant from an estimation
viewpoint.} to a comparator with a random threshold that is
distributed according to $p_X(-x)$. The quantization function
$Q_{it}^{Th}(\cdot)$ in \eqnref{eqn:ourQFunc} can be viewed as a
special case of this where $p_X(-x)$ is the uniform distribution
over $[-c,c]$.  The constructive scheme of \secref{sec:ourscheme}
and the analysis of this work shows that $Q_{it}^{Th}(\cdot)$ can in
fact be used even on noisy field observations with an additive noise
of {\em unknown} distribution.

\subsection{\label{sec:LuoScheme}Parameter Estimation with One--Bit Messages}

The parameter estimation problem in \cite{Luo-IT2005-UDEBCSN}
corresponds to the special case of a spatially constant field
($s(x_i) = s$ for all $i$ where the index $t$ is omitted since
$T=1$) which is addressed by Corollary~\ref{cor:constFieldCase}. We
summarize below the key features of the randomized binary quantizer
proposed in \cite{Luo-IT2005-UDEBCSN} and show that the randomized
$1$--bit SQ function $Q(\cdot)$ of \cite{Luo-IT2005-UDEBCSN} is
statistically and MSE performance--wise equivalent to the uniform
random threshold quantizer $Q_{it}^{Th}(\cdot)$ in
\eqnref{eqn:ourQFunc}. However, the $Q(\cdot)$ function of
\cite{Luo-IT2005-UDEBCSN} implicitly requires sensors of arbitrarily
high precision, a property that is undesirable for sensor hardware
implementations.

In \cite{Luo-IT2005-UDEBCSN}, each sensor $i$ first shifts and
scales it observation $Y_i$ into interval $[0,1]$ creating the value
$\widetilde{Y}_i \triangleq \left(\frac{Y_i + c}{2c}\right)$. Next,
each sensor $i$ generates an auxiliary random variable $\alpha_i$,
which is iid across sensors and is geometrically distributed over
the set of all positive integers: $\pP(\alpha_i = j) = 2^{-j}$ for
all $j \in \{1,2,3,\ldots,\infty\}$. The final quantized binary
sample $B_i$ reported by sensor $i$ is given by the
$\alpha_i^\text{th}$ bit in the binary expansion of
$\widetilde{Y}_i$:
\begin{eqnarray}
B_i &=& Q(Y_i) \triangleq B(\widetilde{Y}_i,\alpha_i), \nonumber \\
&&\mbox{where } \widetilde{Y}_i = \sum_{j=1}^{\infty}
B(\widetilde{Y}_i,j) 2^{-j}. \label{eqn:LuoQFunc}
\end{eqnarray}
Here, $B(\widetilde{Y}_i,j)$ denotes the $j^\text{th}$ bit of
$\widetilde{Y}_i$. For example, if $\widetilde{Y}_i = 0.375$, then
the first four bits of its binary expansion are given by
$B(\widetilde{Y}_i,1) = 0$, $B(\widetilde{Y}_i,2) = 1$,
$B(\widetilde{Y}_i,3) = 1$, and $B(\widetilde{Y}_i,4) = 0$. If
$\alpha_i = 3$, then sensor $i$ reports $B_i = 1$.  This method for
generating binary sensor messages requires sensors to have the
operational ability to quantize an observed real number (the
normalized values $\widetilde{Y}_{i}$) to an arbitrarily high
bit--resolution. Note that the binary values $B_i$ are iid across
all sensors and that its expected value is given by
\begin{eqnarray} \eE[B_i] &=&
\eE_{\widetilde{Y}_i}[\eE_{B_i}[B_i|\widetilde{Y}_i]] \nonumber \\
&=& \eE_{\widetilde{Y}_i}\left[\sum_{j=1}^{\infty} B(\widetilde{Y}_i,j) 2^{-j}\right] \nonumber \\
&=& \eE_{\widetilde{Y}_i}[\widetilde{Y}_i] \nonumber \\
&=& \eE\left[\frac{Y_i + c}{2c}\right] \label{eqn:LuoMsgCondExp} \\
&=& \frac{\eE[s + Z_i] + c}{2c} = \left(\frac{s + c}{2c}\right).
\label{eqn:LuoMsgExp}
\end{eqnarray}

In sharp contrast to the $Q(\cdot)$ function described above, which
requires sensors to have the operational ability to resolve any
arbitrary bit in the binary expansion of their normalized
observations, $Q_{it}^{Th}(\cdot)$ requires only a noisy comparator.
Despite the markedly different operational implementations of
$Q(\cdot)$ and $Q_{it}^{Th}(\cdot)$ (see \eqnref{eqn:LuoQFunc},
\eqnref{eqn:ourQFunc}, and Fig.~\ref{fig:compareImp} which depicts
hardware implementations) they are in fact statistically identical:
the binary quantized values $B_i$ generated by the two schemes have
the same $p_{B_{it}|Y_{it}}(\cdot)$ and $p_{B|s}(\cdot)$ functions
where $p_{B_{it}|Y_{it}}(\cdot)$ is the conditional expectation of
$B_i$ given $Y_i = y_i$ and $p_{B|s}(\cdot)$ is the unconditional
expectation of $B_i$ parameterized by the underlying field value
$s(x_i) = s$. These expectations have been evaluated in
\eqnref{eqn:LuoMsgCondExp}, \eqnref{eqn:LuoMsgExp},
\eqnref{eqn:condExpBinMsg} and \eqnref{eqn:expBinMsg}, and we see
that for both functions
\begin{eqnarray*}
\eE[B_i | Y_i = y_i] = p_{B_{it}|Y_{it}}(y_i) = \left(\frac{y_i +
c}{2c}\right), \text{ and} \\ \eE[B_i] = p_{B|s}(s(x_i)) =
\left(\frac{s(x_i) + c}{2c}\right).
\end{eqnarray*}
This statistical equivalence allows the two quantization functions
$Q(\cdot)$ and $Q_{it}^{Th}(\cdot)$ to be interchanged without
affecting the estimation performance.

\subsection{Extensions to Unbounded Noise and Arbitrary Thresholds with Known Distributions} In this work, we have made assumptions of
zero--mean, amplitude--bounded, additive noise, which can have an
arbitrary, unknown distribution, and uniformly distributed
quantization thresholds. The results of this work can be extended to
deal with noise that is not amplitude--bounded (i.e.  Gaussian,
Laplacian, etc.)  and for thresholds with arbitrary distributions,
however certain technical conditions must be met and the distributions
for both the noise and the threshold must be known.
A possible approach is to combine the noise and threshold random
variables into an overall random dither variable $X_{it} \triangleq
Z_{it} + R_{it}$ and applying the method and results used in
\cite{Masry-IT1981-RASFS} (see \secref{sec:Masry}). The main MSE
result of \thmref{thm:MSEperf} will still hold, however with new
constants multiplying each term in the bound. The method is
essentially the same as in \secref{sec:ourscheme}, however the value
of the field estimate at every point is passed through a non--linear
function, instead of a simple scaling and shifting, given by
\begin{eqnarray*}
g(s) =
\begin{cases}
\mu^{-1}(s) & |s| \leq \mu(a') \\
0 & \mbox{otherwise}
\end{cases}
\end{eqnarray*}
with $\mu(s) \triangleq 1 - 2P_X(-s)$ where $P_X(\cdot)$ is the cdf of
the dither random variable $X_{it}$. The technical requirement for
this extension is that $P_X(\cdot)$ is absolutely continuous with a
probability density function $p_X(\cdot)$ on $(-\infty,\infty)$ which
is continuous and positive over an open interval $(-a',a')$ containing
$[-a,a]$. This ensures that $P_X(\cdot)$ is strictly monotonically
increasing over the signal dynamic range and that $\mu^{-1}(\cdot)$
exists (see \cite{Masry-IT1981-RASFS}).

If the sensor observation noise is unbounded (but the field $s_t(x)$
is still bounded), has zero--mean, and has an {\em unknown} probability
density function (pdf) whose tails decay to zero, it is still possible
to make a weak statement about the achievable MSE as the number of
sensors go to infinity. With unbounded noise, the sensor observations
may exceed any finite dynamic range $[-c,c]$ of the one--bit
sensors. This leads to the appearance of additional error bias terms
(see equation (\ref{eqn:expBinMsg}) and (\ref{eqn:biasResult}) in
Appendix~\ref{app:MSEperfProof}) which depend on the unknown signal
value $s_t(x)$ to be estimated and the dynamic range limit $c$. It can
be shown that these terms go to zero as $c \rightarrow \infty$. Thus
one can assert that for a sufficiently large dynamic range limit $c$
and a corresponding sufficiently large number of sensors $N(c)$, the
MSE can be made smaller than some desired tolerance. The actual
scaling behavior will now also depend on the tail decay rate of the
unknown pdf of the observation noise.

\Section{\label{sec:conclusions}Concluding Remarks}
The results of this work show that for the distributed field
reconstruction problem, for every point of continuity of every field
snapshot, it is possible to drive the MSE to zero with increasing
sensor density while ensuring that the per--sensor bitrate and
sensing--related network overhead rate simultaneously go to zero.
This can be achieved with noisy threshold (one--bit) comparators with
the minimal knowledge of signal and noise dynamic ranges provided that
the noise samples are zero--mean, and independent across sensors and
the underlying field, and the sensor placement and sampling schedule
satisfy a certain uniformity property. The rate of decay of MSE with
increasing sensor density is related to the the local and global
smoothness characteristics of the underlying fields and is
order--optimal for the class of spatially constant fields. This work
has further clarified the utility of randomization for signal
acquisition to combat limited sensing precision and unknown noise
statistics in a distributed sensor network context. This work has also
attempted to systematically account for sensor placement and hardware
issues in addition to the typical constraints encountered in related
studies.

\section*{Acknowledgment} The authors would like to thank Nan Ma and
Manqi Zhao from the ECE department of Boston University for helpful
comments and suggestions during different stages of this work. This
material is based upon work supported by the US National Science
Foundation (NSF) under award (CAREER) CCF--0546598, (CAREER)
ECS--0449194, CCF--0430983, and CNS--0435353, and ONR (PECASE) grant
no.~N00014-02-100362. Any opinions, findings, and conclusions or
recommendations expressed in this material are those of the authors
and do not necessarily reflect the views of the NSF and ONR.

\appendices
\renewcommand{\theequation}{\thesection.\arabic{equation}}
\setcounter{equation}{0}

\renewcommand{\thedefinition}{\Alph{section}.\arabic{definition}}
\renewcommand{\theremark}{\Alph{section}.\arabic{remark}}
\renewcommand{\thetheorem}{\Alph{section}.\arabic{theorem}}

\Section{\label{app:MSEperfProof}Proof of Theorem~\ref{thm:MSEperf}}

First, note that the expected value of the binary message $B_{it}$
is given by
\begin{eqnarray}
\eE[B_{it}] &=& \eE[\mathbf{1}(Y_{it} > R_{it})] \nonumber \\
&=& \eE_{Y_{it}}[\eE_{R_{it}}[\mathbf{1}(Y_{it} > R_{it}) | Y_{it} ]] \nonumber \\
&=& \eE_{Y_{it}}[\pP\left(R_{it} < Y_{it} | Y_{it}\right)] \nonumber \\
&\stackrel{(i)}{=}& \eE_{Y_{it}}\left[\frac{Y_{it} + c}{2c}\right] \label{eqn:condExpBinMsg} \\
&=& \frac{\eE[s_t(x_i) + Z_{it}] + c}{2c} \nonumber \\
&\stackrel{(ii)}{=}& \left(\frac{s_t(x_i) + c}{2c}\right) \label{eqn:expBinMsg},
\end{eqnarray}
which is the value of the field $s_t(\cdot)$ at location $x_i$ shifted
and normalized to the interval $[0,1]$. Note that the key steps are
step $(i)$ where we used the fact that $R_{it}$ is uniformly
distributed over $[-c,c]$ and step $(ii)$ where we used the fact that
$Z_{it}$ has zero mean. It should be noted that the final result
(\ref{eqn:expBinMsg}) holds for any $F_\mathbf{Z}(\mathbf{z}) \in
\mathcal{F}$.

Using \eqnref{eqn:expBinMsg} we can bound the bias and the variance
of the estimator $\widehat{S}_t(x)$. The bound on the MSE follows
from the bounds on these values since, for any estimator of a
non--random parameter, we have
\begin{equation}\label{eqn:MSEidentity}
\mathrm{MSE}\left(\widehat{S}_t(x)\right) =
\mathrm{bias}^2\left(\widehat{S}_t(x)\right) +
\mathrm{var}\left(\widehat{S}_t(x)\right).
\end{equation}

Let $j \in \{1, \ldots, L\}$ denote the index of the supercell that
$x$ falls in. We bound the magnitude of bias of the estimate
$\widehat{S}_t(x)$ in the following way
\begin{eqnarray}
\left|\mathrm{bias}\left(\widehat{S}_t(x)\right)\right| &=&
\left|\eE\left[\widehat{S}_t(x) - s_t(x)\right]\right| \nonumber \\
&=& \Bigg| \eE \Bigg[2c \Bigg[\frac{1}{n} \sum_{i \in I(j,t)} B_{it} \Bigg] \nonumber \\
&& - c - s_t(x) \bigg] \bigg| \nonumber \\
&=& \Bigg| 2c \Bigg[\frac{1}{n} \sum_{i \in I(j,t)} \eE\left[B_{it}\right] \Bigg] \nonumber \\
&&- c - s_t(x) \bigg| \nonumber \\
&\stackrel{(i)}{=}& \Bigg| 2c \Bigg[\frac{1}{n} \sum_{i \in I(j,t)}
\Bigg(\frac{s_t(x_i) + c}{2c}\Bigg) \Bigg] \nonumber \\
&&- c - s_t(x) \bigg| \nonumber \\
&=& \Bigg| \frac{1}{n} \sum_{i \in I(j,t)} \left(s_t(x_i) -
s_t(x)\right) \Bigg| \nonumber \\
&\leq& \frac{1}{n} \sum_{i \in
I(j,t)} \left|s_t(x_i) -
s_t(x)\right| \nonumber \\
&\stackrel{(ii)}{\leq}& \frac{1}{n} \sum_{i \in I(j,t)}
\omega_t \left(\|x - x_i\|,x\right) \nonumber \\
&\stackrel{(iii)}{\leq}& \frac{1}{n} \sum_{i \in I(j,t)}
\omega_t \left(\frac{\sqrt{d}}{\sqrt[d]{L}},x\right) \nonumber \\
&=& \omega_t
\left(\frac{\sqrt{d}}{\sqrt[d]{L}},x\right) \nonumber \\
&\stackrel{(iv)}{\leq}& \widetilde{\omega}_t
\left(\frac{\sqrt{d}}{\sqrt[d]{L}}\right), \label{eqn:biasResult}
\end{eqnarray}
where $(i)$ follows from (\ref{eqn:expBinMsg}), $(ii)$ and $(iv)$
follow from Definitions~\ref{def:localMod} and~\ref{def:globalMod},
and $(iii)$ follows because the local modulus of continuity is a
nondecreasing function of its first argument for each fixed value of
its second argument and since any sensor in the supercell containing
$x$ is within distance $\frac{\sqrt{d}}{\sqrt[d]{L}}$ of $x$ (the
length of the diagonal of a supercell).

The variance of the estimate is bounded by
\begin{eqnarray}
\mathrm{var}[\widehat{S}_t(x)] &=& \mathrm{var}\left[2c
\left[\frac{1}{n} \sum_{i \in I(j,t)} B_{it} \right] - c\right] \nonumber \\
&=& \left(\frac{4c^2}{n^{2}}\right) \sum_{i \in I(j,t)}
\mathrm{var}[B_{it}] \label{eqn:varBoundLine2} \\
&\leq& \left(\frac{4c^2}{n^{2}}\right)\left(\frac{n}{4}\right) =
\left(\frac{LMc^2}{N}\right), \label{eqn:varBoundLine3}
\end{eqnarray}
where we used standard properties of variance and the fact that
$\{B_{it}\}$ are independent to obtain \eqnref{eqn:varBoundLine2},
and we used the fact the variance of a Bernoulli$\{0,1\}$ random
variable is bounded by $(1/4)$ and that $n = (N/(LM))$ to obtain
\eqnref{eqn:varBoundLine3}.

Combining these bounds for the bias and variance given in
\eqnref{eqn:biasResult} and \eqnref{eqn:varBoundLine3} of the
estimator and using the identity in \eqnref{eqn:MSEidentity}, we get
the desired bound on the MSE for all $x \in G$, $t \in \{1, \ldots,
T\}$, and $F_\mathbf{Z}(\mathbf{z}) \in \mathcal{F}$. \proofover

\Section{\label{app:ASConvProof}Proof of Theorem~\ref{thm:ASConv}}

First, we note that
\begin{equation}\label{eqn:ASequiv}
\widehat{S}_t(x) \xrightarrow{\mathrm{a.s.}} s_t(x) \equiv
\left|\widehat{S}_t(x) - s_t(x)\right| \xrightarrow{\mathrm{a.s.}}
0.
\end{equation}
Thus, we proceed with the triangle equality to bound
\begin{eqnarray}\label{eqn:TriangleBoundForAS}
\left|\widehat{S}_t(x) - s_t(x)\right| &\leq& \left|\widehat{S}_t(x)
- \eE\left[\widehat{S}_t(x)\right]\right| \nonumber \\
&+& \left|\eE\left[\widehat{S}_t(x)\right] - s_t(x)\right|.
\end{eqnarray}
In the proof of \thmref{app:MSEperfProof} given in
\appref{app:MSEperfProof} we have shown that the second term of
\eqnref{eqn:TriangleBoundForAS}, which is the absolute value of the
estimator bias, is bounded by \eqnref{eqn:biasResult} which shows
that
\begin{equation}\label{eqn:ASConvSecondTerm}
\left|\eE\left[\widehat{S}_t(x)\right] - s_t(x)\right|
\longrightarrow 0
\end{equation}
as $N$ and $L$ scale as in \eqnref{eqn:LNScaling}.

Letting $j$ denote the supercell that $x$ falls in, the first term
of \eqnref{eqn:TriangleBoundForAS} can be rewritten as
\[
\left|\widehat{S}_t(x) - \eE\left[\widehat{S}_t(x)\right]\right| =
\left| 2c \left[\frac{1}{n} \sum_{i \in I(j,t)} B_{it} - \eE[B_{it}]
\right] \right|.
\]
Recall that the cardinality of $I(j,t)$ is $n = (N/(LM))$. Since the
$B_{it}$ random variables are independent across sensors and their
fourth central moments are uniformly bounded by 1 (since they are
binary~$\{0,1\}$ random variables), the strong law of large numbers
\cite[pp.~206--207]{Durret-TLC-PTE} can be applied to obtain
\[
\frac{1}{n} \sum_{i \in I(j,t)} B_{it} - \eE[B_{it}]
\xrightarrow{\mathrm{a.s.}} 0,
\]
as $N \longrightarrow \infty$ (since $n = (N/(LM))$) and thus the
first term of \eqnref{eqn:TriangleBoundForAS}
\begin{equation}\label{eqn:ASConvFirstTerm}
\left|\widehat{S}_t(x) - \eE\left[\widehat{S}_t(x)\right]\right|
\xrightarrow{\mathrm{a.s.}} 0.
\end{equation}
Combining \eqnref{eqn:ASConvSecondTerm} and
\eqnref{eqn:ASConvFirstTerm} into \eqnref{eqn:ASequiv} and
\eqnref{eqn:TriangleBoundForAS} finishes the proof. \proofover


\bibliography{../LaTeX/biblio}

\begin{thebibliography}{10}
\providecommand{\url}[1]{#1}
\csname url@samestyle\endcsname
\providecommand{\newblock}{\relax}
\providecommand{\bibinfo}[2]{#2}
\providecommand{\BIBentrySTDinterwordspacing}{\spaceskip=0pt\relax}
\providecommand{\BIBentryALTinterwordstretchfactor}{4}
\providecommand{\BIBentryALTinterwordspacing}{\spaceskip=\fontdimen2\font plus
\BIBentryALTinterwordstretchfactor\fontdimen3\font minus
  \fontdimen4\font\relax}
\providecommand{\BIBforeignlanguage}[2]{{%
\expandafter\ifx\csname l@#1\endcsname\relax
\typeout{** WARNING: IEEEtran.bst: No hyphenation pattern has been}%
\typeout{** loaded for the language `#1'. Using the pattern for}%
\typeout{** the default language instead.}%
\else
\language=\csname l@#1\endcsname
\fi
#2}}
\providecommand{\BIBdecl}{\relax}
\BIBdecl

\bibitem{MasryC-IT1981-BPCNT}
{E.~Masry and S.~Cambanis}, ``{Consistent estimation of continuous--time
  signals from nonlinear transformations of noisy samples},'' \emph{IEEE
  Trans.~Info.~Theory}, vol. {IT--27}, pp. {84--96}, {Jan.} 1981.

\bibitem{Masry-IT1981-RASFS}
{E.~Masry}, ``{The reconstruction of analog signals from the sign of their
  noisy samples},'' \emph{IEEE Trans.~Info.~Theory}, vol. {IT--27}, no.~6, pp.
  {735--745}, {Nov.} 1981.

\bibitem{Luo-IT2005-UDEBCSN}
{Z.~Q.~Luo}, ``{Universal decentralized estimation in a bandwidth constrained
  sensor network},'' \emph{IEEE Trans.~Info.~Theory}, vol. {IT--51}, pp.
  {2210--2219}, {Jun.} 2005.

\bibitem{MarcoDLN-IPSN2003-MTCDWSN}
{D.~Marco, E.~J.~Duarte-Melo, M.~Liu, and D.~Neuhoff}, ``{On the many--to--one
  transport capacity of a dense wireless sensor network and the compressibility
  of its data},'' in \emph{{Information Processing in Sensor Networks,
  Proceedings of the Second International Workshop, Palo Alto, CA, USA, April
  22-23, 2003}}, ser. {Lecture Notes in Computer Science edited by L.~J.~Guibas
  and F.~Zhao, Springer, New York, 2003}, {Apr.}, pp. {1--16}.

\bibitem{zorands2000}
{Z.~Cvetkovi\'{c} and I.~Daubechies}, ``{Single bit Oversampled A/D conversion
  with Exponential accuracy in bit rate},'' \emph{{Proc.~Data Compression
  Conference}}, pp. {343--352}, {Mar.~} 2000.

\bibitem{zorandli2002}
Z.~Cvetkovi\'{c}, I.~Daubechies, and B.~F. Logan, ``{Interpolation of
  Bandlimited functions from quantized Irregular Samples},'' \emph{{Proc.~Data
  Compression Conference}}, pp. {412--421}, {Apr.~} 2002.

\bibitem{IshwarKR-2003-DSDSNBCP}
{P.~Ishwar, A.~Kumar, and K.~Ramchandran}, ``{Distributed sampling for dense
  sensor networks: A ``bit-conservation'' principle},'' in \emph{{Information
  Processing in Sensor Networks, Proceedings of the Second International
  Workshop, Palo Alto, CA, USA, April 22-23, 2003}}, ser. {Lecture Notes in
  Computer Science edited by L.~J.~Guibas and F.~Zhao, Springer, New York,
  2003}, pp. {17--31}.

\bibitem{KumarIR-2004-DSSNBLF}
{A.~Kumar, P.~Ishwar, and K.~Ramchandran}, ``{On distributed sampling of smooth
  non-bandlimited fields},'' in \emph{{Proc.~Third Inttl.~Symposium Information
  Processing in Sensor Networks}}.\hskip 1em plus 0.5em minus 0.4em\relax {New
  York, NY}: {ACM Press}, 2004, pp. {89--98}.

\bibitem{BergerZV-IT1996-TCP}
{T.~Berger, Z.~Zhang, and H.~Viswanathan}, ``{The CEO problem [multiterminal
  source coding]},'' \emph{IEEE Trans.~Info.~Theory}, vol. {IT--42}, pp.
  {887--902}, {May.} 1996.

\bibitem{ViswanathanB-IT1997-QGCP}
{H.~Viswanathan and T.~Berger}, ``{The quadratic Gaussian CEO problem},''
  \emph{IEEE Trans.~Info.~Theory}, vol. {IT--43}, pp. {1549--1559}, {Sep.}
  1997.

\bibitem{PrabhakaranTR-ISIT04-RQGCP}
{V.~Prabhakaran, D.~Tse, and K.~Ramchandran}, ``{Rate-region of the quadratic
  Gaussian CEO problem},'' in \emph{{Proc.~IEEE International Symposium on
  Information Theory}}, {Chicago, IL}, {Jun.} 2004, p. 119.

\bibitem{KashyapLXL-2005-DSCDSN}
{A.~Kashyap, L.~A.~Lastras-Montano, C.~Xia, and Z.~Liu}, ``{Distributed source
  coding in dense sensor networks},'' in \emph{{Proc.~Data Compression
  Conference}}, {Snowbird, UT}, {Mar.} 2005, pp. {13--22}.

\bibitem{NeuhoffP-2006-UPRIDLSC}
{D.~L.~Neuhoff and S.~S.~Pradhan}, ``{An upper bound to the rate of ideal
  distributed lossy source coding of densely sampled data},'' in
  \emph{{Proc.~IEEE International Conference on Acoustics, Speech and Signal
  Processing}}, {Toulouse, France}, {May} 2006, pp. {1137--1140}.

\bibitem{Cvetkovic-IT2003-RPREUANQ}
{Z.~Cvetkovi\'{c}}, ``{Resilience properties of redundant expansions under
  additive noise and quantization},'' \emph{IEEE Trans.~Info.~Theory}, vol.
  {IT--49}, pp. {644--656}, {Mar.} 2003.

\bibitem{GastparV-2003-SCCSN}
{M.~Gastpar and M.~Vetterli}, ``{Source--channel communication in sensor
  networks},'' \emph{{Lecture Notes in Computer Science}}, vol. 2634, pp.
  {162--177}, {Apr.} 2003.

\bibitem{GastparRV-2003-TCNCLSCCR}
{M.~Gastpar, B.~Rimoldi, and M.~Vetterli}, ``{To Code, or not to code: Lossy
  source--channel communication revisited},'' \emph{IEEE Trans.~Info.~Theory},
  vol. {IT--49}, pp. {1147--1158}, May 2003.

\bibitem{NowakMW-2004-EIFWSN}
{R.~Nowak, U.~Mitra, and R.~Willet}, ``{Estimating inhomogenous fields using
  wireless sensor networks},'' \emph{IEEE J.~Sel.~Areas Commun.}, vol.~22,
  no.~6, pp. {999--1006}, {Aug.} 2004.

\bibitem{LiuES-2005-OPFESN}
{K.~Liu, H.~El--Gamal, and A.~Sayeed}, ``{On optimal parametric field
  estimation in sensor networks},'' in \emph{{Proc.~IEEE/SP 13th Workshop on
  Statistical Signal Processing}}, {Jul.} 2005, pp. {1170--1175}.

\bibitem{BajwaSN-2005-MSCCFEWSN}
{W.~Bajwa, A.~Sayeed, and R.~Nowak}, ``{Matched source--channel communication
  for field estimation in wireless sensor networks},'' in \emph{{Proc.~Fourth
  Inttl.~Symposium Information Processing in Sensor Networks}}, {Apr.} 2005,
  pp. {332--339}.

\bibitem{LiuU-2006-ODPTGSN}
{N.~Liu and S.~Ulukus}, ``{Optimal distortion--power tradeoffs in Gaussian
  sensor networks},'' in \emph{{Proc.~IEEE International Symposium on
  Information Theory}}, {Seattle, WA, USA}, {Jul.} 2006, pp. {1534--1538}.

\bibitem{ZhaoST-2006-IBSPNSN}
{Q.~Zhao, A.~Swami, and L.~Tong}, ``{The interplay between signal processing
  and networking in sensor networks},'' \emph{{IEEE Signal Processing
  Magazine}}, vol.~23, no.~4, pp. {84--93}, {Jul.} 2006.

\bibitem{AliprantisB-AP90-PRA}
{C.~D.~Aliprantis and O.~Burkinshaw}, \emph{{Principles of Real
  Analysis}}.\hskip 1em plus 0.5em minus 0.4em\relax {San Diego, CA}: {Academic
  Press}, 1990.

\bibitem{ChouPR-Asilomar2002-TECDSN}
{J.~Chou, D.~Petrovic, and K.~Ramchandran}, ``{Tracking and exploiting
  correlations in dense sensor networks},'' in \emph{{Proc.~Annual Asilomar
  Conference on Signals, Systems, and Computers}}, {Pacific Grove, CA}, {Nov.}
  2002.

\bibitem{KahnKP-MOBICOM99-NCCMNSD}
\BIBentryALTinterwordspacing
{J.~M.~Kahn, R.~H.~Katz, and K.~S.~J.~Pister}, ``{Next century challenges:
  Mobile networking for ``Smart Dust''},'' in \emph{{Proc.~ACM International
  Conference on Mobile Computing and Networking ({MOBICOM})}}, {Seattle, WA},
  {Aug.} 1999, pp. {271--278}. [Online]. Available:
  \url{{citeseer.nj.nec.com/kahn99next.html}}
\BIBentrySTDinterwordspacing

\bibitem{CoverJ-1991-EoIT}
{T.~M.~Cover and J.~A.~Thomas}, \emph{{Elements of Information Theory}},
  1st~ed.\hskip 1em plus 0.5em minus 0.4em\relax {New York, NY}:
  {Wiley--Interscience}, 1991.

\bibitem{Kay-1993-FSSPET}
{S.~M.~Kay}, \emph{{Fundamentals of Statistical Signal Processing, Volume I:
  Estimation Theory}}, 1st~ed.\hskip 1em plus 0.5em minus 0.4em\relax {Upper
  Saddle River, NJ}: {Prentice--Hall}, 1993, vol.~1.

\bibitem{Durret-TLC-PTE}
R.~Durret, \emph{Probability: Theory and Examples}.\hskip 1em plus 0.5em minus
  0.4em\relax Thomson Learning College, 1990.

\end{thebibliography}


\end{document}